\documentclass[12pt,preprint]{aastex}

\shorttitle{WD \& SNeIa}
\shortauthors{Ablimit at el.}

\begin{document}


\title{Possible Contribution of Magnetized White Dwarf Binaries to Type Ia Supernova Populations}
\author{Iminhaji Ablimit\altaffilmark{1, \star} and Keiichi Maeda\altaffilmark{1}}
\altaffiltext{}{Corresponding author: Iminhaji Ablimit (iminhaji@kusastro.kyoto-u.ac.jp)}
\altaffiltext{1}{Department of Astronomy, Kyoto University, Kitashirakawa-Oiwake-cho, Sakyo-ku, Kyoto 606-8502, Japan}
\altaffiltext{}{$^\star$JSPS International Research Fellow }


\begin{abstract}

The evolution of an accreting white dwarf (WD) with strong magnetic field toward a type Ia supernova (SN Ia) may differ from the classical single-degenerate (SD) channel. In this paper, we perform binary population synthesis (BPS) simulations for the SD channel with a main-sequence (MS) companion, including the strongly magnetized WD accretion. Under a reasonable assumption that the fraction of such systems is $\sim 15$\%, the resulting delay time distribution (DTD) roughly follows the $t^{-1}$ power-law distribution. Within the (WD/MS) SD channel, the contribution from the highly magnetized WD is estimated to be comparable to that from the classical, non-magnetized WD channel. The contribution of the SD channel toward SNe Ia can be at least $\sim 30$\% among the whole SN Ia population. We suggest that the SNe Ia resulting from the highly-magnetized WD systems would not share observational properties expected for the classical SD channel; for every (potentially peculiar) SN observationally associated with the SD channel, we expect a comparable number of the `hidden' SD population in the normal class. 

\end{abstract}


\section{Introduction}

The type Ia supernova (SN Ia) is an excellent distance indicator used in the cosmological measurement, as well as an important iron contributor in the chemical evolution (Greggio \& Renzini 1983; Riess et al. 1998). Despite the importance, the origin of SNe Ia is largely obscured (Hillebrandt \& Niemeyer 2000; Maeda \& Terada 2016; Livio \& Mazzali 2018). The scenario involving a thermonuclear explosion of a carbon-oxygen white dwarf (CO WD) in a close binary system has been widely accepted (Holey \& Folwer 1960; Tutukov \& Yungelson 1979; Webbink 1984). Two scenarios have been discussed as the main channel(s) toward SNe Ia. In the double degenerate (DD) channel, two WDs merge by energy and angular momentum loss due to the gravitational wave radiation (Webbink 1984; Iben \& Tutukov 1984). In the single degenerate (SD) channel, the WD accretes matter from its non-degenerate companion star and grows in mass to the Chandrasekhar mass (Whelan \& Iben 1973; Nomoto 1982; Di Stefano et al. 2011).

The SN Ia birth rate and delay time (from the star formation to the SN Ia explosion) distribution (DTD) place strong constraints on the SN Ia progenitor evolution (e.g., Wang \& Han 2012; Maoz, Mannuccin \& Nelemans 2014). The observationally inferred DTD indicates a power-law function with an index of -1 (Maoz \& Mannucci 2012), and it spans from the `prompt' (young) component with delay times less than 0.1 Gyr (Mannucci et al. 2006; Aubourg et al. 2008) and the `delayed' component with delay times $\ge 2-3$ Gyr (Botticella et al. 2008; Totani et al. 2008). Schawinski (2009) analyzed 21 nearby SNe Ia in early-type galaxies, and demonstrated a lack of young SNe Ia in these early-type galaxies by deriving the minimum delay times from 0.28 Gyr to 1.25 Gyr in their SNe Ia sample. 
 
The DD channel has been argued to naturally produce the power-law DTD. On the other hand, it has been sometimes regarded as a challenge for the SD channel. There are a few characteristic evolutionary timescales in the SD scenario set by the nature of the donor/companion stars; main-sequence (MS), red giant (RG) or even He stars. Generally, the MS system and the RG system are suggested to be responsible for the (relatively) young and old populations, respectively (Kobayashi \& Nomoto 2009). The He donor channel has been suggested to create the young(-est) (or `prompt') population (Wang et al. 2009; Claeys et al. 2014). As such, it might not be straightforward to explain the power-law DTD as a combination of these different evolutionary channels within the SD scenario (but see Kobayashi et al. (2019)). 

Indeed, the young SN Ia population is generally missing in most progenitor evolution models (Maoz, Mannuccin \& Nelemans 2014; Ablimit, Maeda \& Li 2016; Ilkov \& Soker 2011; Soker 2019); this does not seem to be reconciled by various uncertainties involved in the binary population synthesis (BPS) simulations (Jorgensen et al. 1997; Yungelson \& Livio 2000; Wang et al. 2009; Ablimit et al. 2016). The SD (with the H-rich donor) generally predicts the delay times longer than $\sim$ 1 Gyr (e.g., L$\ddot{\rm u}$ et al. 2009; Wang \& Han 2012; Claeys et al. 2014). This might be an importance constraint on the contribution of the SD channel within the whole SN Ia populations. Assuming a combination of the SD channel and another channel, with the latter having the power-law DTD (e.g., the prediction by the DD scenario), if the fraction of the SD would be large, then this lack of the young population should be reflected to the combined DTD. This would create a tension to the observationally inferred DTD. Adding the He donor contribution would be a solution, but this might require a fine tuning on the relative contributions from the H-rich donors and He-rich donors. 

Given the diversities seen in observational properties of SNe Ia, it is likely that multiple progenitor channels are realized in nature. There have been observational indications that a few individual classes of peculiar SNe are associated with the SD scenario (\S 3.3; see also Maeda \& Terada 2016). It is thus important to understand the contribution by the SD channel to the normal and whole populations of SNe Ia. The WD+MS channel is one of the important contributors among the SD scenarios. The WD+MS channel has three different evolution pathways (see Wang \& Han (2012) for more details). (1) When the primordial primary is in the Hertzsprung gap or first giant branch stage, it can fill its Roche lobe and the two stars may evolve into the common envelope (CE) due to a large mass ratio or a convective envelope of the mass donor star. After the CE ejection and end of He burning of the primary, the CO WD can be formed with the MS companion. This channel is realized for the following ranges in the initial binary parameters; the initial primary mass $M_{1,\rm{i}} \sim 4.0-7.0M_\sun$, the initial mass ratio $q=M_{2,\rm{i}}/M_{1,\rm{i}}\sim0.3-0.4$ and the initial orbital period $\rm{log} P \sim 1.0-1.5$ in days. (2) When the primary evolves to the early asymptotic giant branch, a CE may be formed due to the dynamically unstable mass transfer. After the CE is expelled, the primary finally becomes a CO WD with the secondary MS. This channel is realized for the following initial parameters; $M_{1,\rm{i}} \sim 2.5-6.5M_\sun$, $q=M_{2,\rm{i}}/M_{1,\rm{i}}\sim0.2-0.9$ and $\rm{log} P \sim 2.0-3.0$. (3) The mass transfer could be dynamically unstable and a CE could be formed when the primary evolves to the the thermal pulsing asymptotic giant branch. After the CE ejection, a CO WD + MS system is produced. This channel is for the initial parameters: $M_{1,\rm{i}} \sim 3.0-6.5M_\sun$, $q=M_{2,\rm{i}}/M_{1,\rm{i}}\sim0.2-0.7$ and $\rm{log} P \sim 2.5-3.5$ (e.g., Wang 2018). The first two evolution pathways are the main routes to the WD+MS channel in the SD model for the SNe Ia, and we include all of them to obtain the SNe Ia rate from the WD+MS channel.

One key physical process which might have been missing in the context of the SD scenario is the accretion under strong magnetic field (e.g., Wheeler 2012; Ablimit et al. 2014; Farihi et al. 2017). About 25\% of Cataclysmic variables (CVs) are magnetic CVs (Ferrario et al 2015). The fraction can indeed be even higher ($\sim$33\%) according to the recent result from GAIA DR2 (Pala et al. 2019). Magnetic WDs in supersoft X-ray sources and symbiotic binaries have been discovered as well (Kahabka 1995; Sokoloski \& Bildsten 1999; Osborne et al. 2001). Ablimit \& Maeda (2019) studied the evolution of the magnetized WD binary toward the SN Ia under the magnetic confinement accretion model. In this paper, we investigate the possible contribution of the magnetized WD/MS binaries in the SN Ia population with BPS simulations. In \S 2, method and models are described. The DTD and birth rate based on our models are given in \S 3. We further address possible contribution of the magnetized WD accretion channel to the SN Ia progenitors and implications for the SN Ia diversities. Our findings are summarized in \S 4.

\section{Method}
\label{sec:model}

\subsection{The magnetic confinement model}

In the polar-like systems, the binary likely has a synchronous rotation without accretion disk formation (Cropper 1990). The stream of matter from the donor is captured by the magnetic field of the WD, then it follows the magnetic field lines and falls down onto the magnetic poles of the WD as an accretion column. This may affect the accretion process onto a WD and the condition for nova eruptions (Livio 1983), and may further affect the mass growth of a WD toward the Chandrasekhar mass ($M_{\rm Ch}$) to form an SN Ia progenitor (Ablimit \& Maeda 2019). Previous studies suggested a possibility of producing the super-Chandrasekhar mass (above 2 $M_\sun$) WDs (e.g., Das et al. 2013) supported by the strong magnetic field in the interior of a WD. However, such an effect is significant only for an extremely strong magnetic field ($10^{15} \sim 10^{17}$ G), which far exceeds the conventionally assumed values ($10^{8} \sim 10^{10}$ G, see Shapiro \& Teukolsky 1983). Thus, we adopt the classical Chandrasekhar mass ($M_{\rm Ch}=1.38M_\sun$) in our study.
 

Once the magnetic column accretion takes place, it increases the density of the accreting material for a given accretion rate. Namely, the WD surface would feel locally as if the accretion rate would be higher than the mass transfer rate from the secondary. Thanks to this increase in the local/effective accretion rate, it can accrete the material steadily for a low transfer rate that would instead lead to a nova eruption without the magnetic confinement. Ablimit \& Maeda (2019) studied the WD $+$ MS star binary evolution taking into account this effect. They found that (1) the lower limit for the initial mass of the donor star, leading to the WD mass increase to $M_{\rm Ch}$, can be lower than those found in previous works (e.g., Li \& van den Heuvel 1997), (2) the range of the initial orbital period of the magnetized WD binary leading to the Chandrasekhar mass WD is extended,  covering from 0.3 to 25 days. The resulting parameter space for the SN Ia progenitors is substantially larger than those found in previous work (see Ablimit \& Maeda (2019) for details).

\subsection{Binary population synthesis}

We perform a series of Monte-Carlo simulations for the WD/MS binary evolution, by using the BSE (binary stellar evolution) population synthesis code (Hurley et al. 2002).  In this binary population synthesis (BPS) calculation, $10^{7}$ binary systems are followed from the initial MS/MS configuration through their evolutions. The systems which end up with the formation of the Chandrasekhar mass WD through the accretion from a secondary MS are tagged as the SN Ia progenitors. Note that we have not considered WD/RG (red giant) systems in this study. Note also that the BPS code follows the evolution of individual binary systems explicitly, and the modification has been made in the treatment of the mass accretion efficiency for the highly magnetized WD according to Ablimit \& Maeda (2019, see the section 2 of the paper). Our recipe to compute the mass transfer is the following (see Ablimit \& Maeda (2019) for more details):  The H burning efficiency (as well as the He burning efficiency) is determined by the RLOF mass transfer $\dot{M}$ in the classical model without the magnetic field.  For the highly magnetized WD, we replace $\dot{M}$ by the polar mass transfer ($\dot{M_{\rm p}}$ ) in computing the burning (and thus accretion) efficiency. This efficiency is multiplied to the mass transfer ($\dot{M}$) to obtain the mass accretion rate to the primary WD.

The initial input parameters of the primordial binaries are set as follows. The initial mass function of Kroupa et al. (1993) is adopted for the masses of the primary stars. The masses of the secondary stars are determined by the distribution of the mass ratio of the secondary to the primary, which is set to be flat between 0 and 1. All stars are assumed to be initially in binaries. The distribution of the orbital separations is assumed to be flat in logarithm. The uniform (flat) initial eccentricity distribution is assumed in a range between 0 and 1. We set the initial metallicity of stars to be the solar value. In calculating the Galactic SN rate, we simply assume a constant star formation rate of $5 M_\sun \ \rm{yr^{-1}}$ (Willems \& Kolb 2004). For the other parameters, the default values adopted in the original BSE population synthesis code (Hurley et al. 2002) are used in this paper. 
Given the total mass $M_{total}$ in the simulation and the initial sets of $\phi(M_{\rm 1}, M_{\rm 2}, a)$
(where $M_{\rm 1}$, $M_{\rm 2}$ and $a$ are the primary star mass, companion star mass and binary separation, and $\phi$ is the distributions of them), the DTD is computed as follows,
\begin{equation}
{\rm{DTD}}(t) \sim \int_{M_{\rm 1}} \int_{M_{\rm 2}} \int_{a} \frac{1}{M_{total} \Delta{t}}\delta({\rm{SN}\,I}a) \phi(M_{\rm 1}, M_{\rm 2}, a) dM_{\rm 1} dM_{\rm 2} da.
\end{equation} 
If the binary does not make an SN Ia, then $\delta({\rm{SN}\,I}a) = 0$. When the binary produces an SN Ia during a time interval between $t$ and $t+\Delta{t}$, $\delta({\rm{SN}\,I}a) = 1$.

Details of our BPS code have been described by Ablimit, Maeda \& Li (2016) and Ablimit \& Maeda (2018). As compared to the original code by Hurley et al. (2002), we have updated the treatment of the mass transfer and the common envelope (CE) evolution. As in other BPS works, the fate of the CE phase is determined by the CE ejection efficiency ($\alpha_{\rm CE}$) and the binding energy parameter ($\lambda$). The CE efficiency, $\alpha_{\rm CE}$, is taken to be 1.0 in this paper. For the binding energy parameter, we use $\lambda = \lambda_{\rm e}$ (which includes the contributions of the gravitational energy, internal energy and entropy of the envelope; Wang et al. 2016). Note that only the gravitation energy is counted in many BPS works ($\lambda = \lambda_{\rm g}$), and we will briefly address how this affects the outcome of the binary evolution. In this paper, we fix the BPS parameters (such as the mass ratio distribution) adopting typical values frequently used for these parameters; our main aim in this paper is to demonstrate the potential importance of the highly magnetized magnetic WDs toward SNe Ia. There are intensive studies in the general BPS framework on the rate of SNe Ia, on how the BPS parameters affect the outcomes (e.g., Claeys et al. 2014; Ablimit et al. 2016). Similarly, we have adopted a very simple star-formation rate in our study, for the same reason.

We focus two models in the BPS calculations, named Model non-B and Model high-B. All WD/MS binaries are assumed to be non-magnetic WD binaries in Model non-B. In discussing the shape of the DTD, we assume all WDs are born with the high magnetic field in the WD/MS binaries in Model high-B, while the other parameters are set to be the same with Model non-B. In order to clearly show the possible contribution of the magnetic confinement model, we adopt same typical initial conditions in BPS calculations.

When we discuss the contribution of Model high-B to the total SN rate as combined with Model non-B, we will adopt $15$\% as the fraction of the highly magnetized WDs. Around 25\% of WDs have the magnetic field among the known CVs, and about 15\% -- 18\% are polars based on the observational results of Ferrario et al (2015; see also Kepler et al. 2013, 2015). Pala et al. (2019) analyzed a sample of 42 CVs within 150 pc based on the GAIA DR2 survey data, and they found that the fraction of magnetic CVs in the volume-limited sample is remarkably high (33\%). In their sample, the fraction of polars is $\sim$28\%. These studies suggest that the evolution of magnetic systems has to be included in the WD binary population models. We conservatively assume that the polar-like systems among the WD binary population is 15\%. When we discuss the contribution of the WD/MS SD channel to the SN Ia populations, we thus combine Model non-B and Model high-B with the fractions of $85$\% and $15$\%, respectively.

\section{Results}

\subsection{The delay time distribution}

The distribution of the time interval between the star formation and SNe Ia explosion (i.e., DTD) is one basic feature characterizing the nature of the SNe Ia progenitors. Observationally, the DTD can be obtained by associating the rate of SNe Ia to the age of the host galaxies (Totani et al. 2008; Maoz \& Mannucci 2012) or that of the local SN site (Maoz \& Badenes 2010). Figure 1 shows the observationally inferred DTD (Maoz \& Mannucci 2012).  
We can extract the DTD in the BPS simulation; it is the SN rate versus the time passed from a brief burst of star formation that formed a unit of stellar mass. For a given stellar mass formed by a single burst, we count the number of systems in which the primary WD reaches to $M_{\rm Ch}$ with the delay time between $t$ and $t + \Delta{t}$ (see \S 2.2).

Figure 1 shows the DTD based on our models. We first briefly address the effect of the CE prescription, namely the binding energy parameter; it is not a main focus of this paper, but this affects the outcome of our reference Model non-B. For the same Model non-B, we simulate an additional model where we set $\lambda = \lambda_{\rm g}$ (which includes only the gravitational potential energy). The delay times of SNe Ia from this model ($\lambda{\rm g}$) are distributed from $\sim 0.3$ and $\sim 3.1$ Gyr. SNe Ia with relatively long delay times are realized with $\lambda=\lambda_{\rm e}$. Because $\lambda_{\rm e}$ leads to a higher efficiency in ejecting the CE, the binaries can survive the CE more easily (Ablimit et al. 2016). While this effect results in a relatively large SN rate in Model non-B, the contribution of DTDs from the non-B model are still largely consistent with previous BPS works (see \S 1).

The young SN Ia population with the delay time less than $\sim$0.1 Gyr is missing in Model non-B, as is consistent with the previous works (e.g., Wang \& Han 2012; Claeys et al. 2014). Model high-B results in a larger range of delay times (between $\sim 0.06$ and $\sim 11$ Gyr). The young and old populations found in Model high-B both originate from the systems with a low mass transfer rate, which would instead experience nova eruptions (and not evolve to $M_{\rm Ch}$) in the absence of a strong magnetic field. Consider a case where a massive primary (with the initial mass above $\sim 5 M_\odot$) evolves to a massive WD ($\sim 1.2 M_\odot$) within 0.1 Gyr and the system becomes a close binary through the CE phase. The mass transfer rate in such a system is generally low and thus would not lead to the SN Ia explosion in the classical model. In Model high-B, with the mass transfer rate of $\sim 10^{-8} M_\odot$ yr$^{-1}$, the WD can evolve to $M_{\rm Ch}$ in a few $\times 10^{7}$ yr. This evolutionary path leads to the young SN Ia population in Model high-B. The old population can arise in a similar manner, but with a lower mass primary star and/or a lower mass secondary star, and/or a longer period which all delay the evolutionary time scale. The examples of the old population can be found in Figures 2 \& 3 of Ablimit \& Maeda (2019). 

It is seen in Figure 1 that the contribution of Model high-B is substantial to shape the DTD, even if the fraction of Model high-B is reduced to $15$\% of the total WD/MS systems. Indeed, the predicted DTD roughly follows the $t^{-1}$ dependence in a large range of the delay time, as is similar to the observationally inferred DTD, even though the predicted rate as a combination of Model non-B and high-B is still short by a factor of a few (see \S 3.2). If there would be another evolutionary pathway which shows the similar $t^{-1}$ dependence (e.g., the DD channel), the combination of such a population to the WD/MS SD channel in this study is still expected to create the $t^{-1}$ dependence. Namely, the SD channel proposed in this paper can contribute to $\sim 50$\% of the whole SN Ia population without a major tension in the resulting DTD, as long as the population with the delay time less than $\sim 10$ Gyr is concerned. Note that we have not included the WD/RG channel, but it would not contribute such a (relatively) young population.

\subsection{The birth rate}

The birth rate of SNe Ia in the Milky Way galaxy ($\sim 3\times10^{-3}\,\rm{yr}^{-1}$; Capprllaro \& Turatto 1997) is one important constraint on the SNe Ia progenitor models. Figure 2 shows the Galactic birth rate of the SNe Ia found in our models. Model non-B results in the Galactic birth rate of $\sim 6.8 \times 10^{-4} (1-f_{\rm B})$ yr$^{-1}$, where $f_{\rm B}$ is the fraction of the highly magnetized WD in the accreting systems. As a reference, Model non-B with $\lambda_{\rm g}$ for the binding energy parameter results in $\sim 3.1 \times 10^{-4} (1-f_{\rm B})$ yr$^{-1}$. The calculated rate from the WD/MS channel without the magnetic field confinement is consistent with a range of the birth rate derived by previous works (Wang \& Han 2012; Claeys et al. 2014).

The rate found in Model high-B is $\sim 3.3\times10^{-3} f_{\rm B}$ yr$^{-1}$. Assuming the fiducial value, $f_{\rm B} = 0.15$ as combined with Model non-B (with 85\% contribution), the predicted total rate is $\sim 1.1 \times 10^{-3}$ yr$^{-1}$. The contributions from Model non-B and Model high-B are comparable, i.e. roughly $\sim 5 \times 10^{-4}$ yr$^{-1}$ for each. The rate of the `young' population within Model high-B is $\sim 5 \times 10^{-5}$ yr$^{-1}$, and it is also the case for the `old' population from Model high-B.  

The total SN Ia rate through the WD/MS channel studied in this paper ($\sim 1.1 \times 10^{-3}$ yr$^{-1}$) can thus account for one third of the Galactic SN Ia rate ($\sim 10^{-3}$ yr$^{-1}$). Given that we omit the WD/RG channel (therefore the old population) in this study, the SD channel could provide a major contribution to the SN Ia rate. A sum of the `young' and `old' populations arising from Model high-B have contributions of $\sim 10$\% to the total SN Ia rate in the WD/MS SD channel (as a combination of Model non-B and high-B), or $\sim 3$\% of the observationally inferred Galactic SN Ia rate. The remaining fraction accounts for the `intermediate' population, which dominates the SNe Ia in the present study (note that all SNe Ia from Model non-B belong to the intermediate population). Note however that these fractions depend on the definition of the `young' and `old' populations, and thus these values should be taken merely as a rough estimate. Further, the value here for the `old' population should not be directly compared to any observational indication, as we do not include the WD/RG channel.

\subsection{Implications for sub-classes of SNe Ia}

While the SN Ia progenitors in Model high-B evolve through the WD/MS SD channel, the nature of the progenitor systems may be quite different from the classical SD channel (i.e., Model non-B). Below, we first summarize the expected nature of the progenitor systems and the SN itself evolved through Model high-B. Possible relations to observed sub-classes of SNe Ia will then be discussed. 

Figure 3 shows distributions of final masses of companion stars just before the SN explosion. Overall, Model high-B is characterized by a lack of a massive MS companion star with $> 1.8 M_\odot$. Because the high mass stars drive the system into the CE in Model high-B, the massive companion stars are missing. Therefore, the detection of the companion star, either in pre-SN images of extragalactic SNe or within SN remnants, is more difficult than Model non-B. Generally, the younger population, in terms of the delay time, tends to have a more massive companion star. For the old population with the delay time exceeding $\sim 4$ Gyr, the companion stars are all less massive than $\sim 0.8 M_\odot$, down to $\sim 0.3 M_\odot$. 

There have been a few reports about non-detection of the companion stars (Li et al. 2011, Kelly et al. 2014, Gonz$\acute{\rm a}$lez Hern$\acute{\rm a}$ndez et al. 2012), but there is only one case where the upper limit could be sufficiently deep to probe a companion star in Model high-B (Schafer \& Pagnotta 2012). Even in this case, the companion star expected in the `old' population (but in the WD/MS channel) would not have been detected. 

Figure 3 also shows the distribution of the `solid angle' of a companion as viewed from the SN ejecta, i.e., the ratio of the companion radius to the binary separation. Note that we consider only the WD/MS channel and do not include the WD/RG systems toward SNe Ia. The donors in our simulations experience either the Case A or early Case B RLOF mass transfer. It has been suggested that the hydrodynamical interaction between the expanding SN ejecta and a companion's envelope should provide an additional heat to the SN ejecta, which is then lost as a blue UV/optical emission in a few days (Kasen 2010). The solid angle is a measure of the strength and frequency of such a bright precursor. As shown in Figure 3, companion stars in Model high-B are substantially more compact relative to the binary separation than in Model non-B. Therefore, we do not expect a strong signal of this companion interaction in SNe Ia evolved through Model high-B. Similarly, the amount of the hydrogen-rich companion envelope stripped into the SN ejecta (Marietta et al. 2000) will be smaller in Model high-B; the non-detection of Balmer series in late-time spectra of normal SNe Ia has been an argument against the SD channel (Mattila et al. 2005; Maeda et al. 2014; Tucker et al. 2019), but this constraint would also be weakened for Model high-B. 

Observationally, SNe Ia showing the possible early blue enhancement are relatively rare. The bright SNe 1991T/1999aa-like SNe Ia tend to show such an emission (Stritzinger et al. 2018). However, it has been suggested that this would likely originate in a different mechanism, e.g., a mixing of $^{56}$Ni out to the surface (Maeda et a. 2018). This is further complicated since there is another mechanism, the He detonation on the surface, which could result in the signature similar to the companion interaction (Jiang et al. 2017; Noebauer et al. 2017; Maeda et al. 2018; De et al. 2019). So far, there is only a single SN Ia where the early blue emission has been most likely attributed to the companion interaction (Cao et al. 2015), indicating an existence of the SD (but non-B) channel. The prediction from model high-B could be largely consistent with the reality of such signals in normal SNe Ia. 

The nature of the progenitor WD may also be different between Model high-B and (classical) Model non-B. In the classical SD model, the disk accretion would supply angular momentum to the WD, and the WD spin may likely be kept high. The WD may even evolve toward the Chandrasekhar mass with the critical rotation (Uenishi et al. 2003; Yoon \& Langer 2004). This may introduce a diversity in the angular momentum, and perhaps even the mass, of the progenitor WD. Further, this may delay the explosion of the WD by the spin-down time scale (Hachisu et al. 2012), again introducing uncertainty and possible diversity to the progenitor WD system. On the other hand, the spin of the WD progenitor in Model high-B may be negligible, as a large fraction of the WD glow through the polar accretion. As such, the nature of the SN Ia progenitors in Model high-B is likely more uniform than in model non-B. Therefore, the nature of the progenitor through Model high-B may form a relatively uniform system as compared to Model non-B. 

Since the WDs in Model high-B tend to evolve with a low mass transfer rate, the mass budget that can be potentially ejected to the circumstellar environment is also small. Therefore, we do not expect the existence of massive and dense CSM around the SN Ia progenitors. The mass transfer rate is typically $\sim 10^{-8} M_\odot$ yr$^{-1}$, and the fraction of the mass ejected to the environment is expected to be small. As such, the system could easily accommodate the upper limit on the CSM density derived for some (normal) SNe Ia (Chomiuk et al. 2012; Margutti et al. 2012; Maeda et al. 2015). 

Overall, the natures of the progenitors and explosions expected in model high-B may indeed have similarities to those expected in the DD channel, rather than the classical SD channel. Recently, there are indications that some outliers may be related to the SD channel, while normal SNe Ia tend not to show any signatures expected in the (classical) SD scenario in terms of the companion and the CSM environment (see Maeda \& Terada 2016 for a review). 

SNe Ia which have been suggested to be linked to the SD channel include the following classes; bright SN 1991T/1999aa-like SNe Ia, over-luminous SNe Ia (i.e., `super-Chandrasekhar candidates'), and SNe Iax. SN 1991T/1999aa-like SNe Ia are associated with a class of `SNe Ia-CSM' (Dilday et al. 2012; Leloudas et al. 2015), which show a distinct signature of massive CSM around the progenitor system. Similarly, at least one of the over-luminous SNe Ia (SN 2012dn) shows a clear signature of dusty CS environment (Yamanaka et al. 2016; Nagao, Maeda \& Yamanaka 2017). These indications of massive and dense CSM suggest that at least a fraction of these bright SNe Ia are better explained by the (classical) SD channel. The same argument then applied to conclude that they are not likely products through Model high-B. 

Another peculiar class of SNe Iax (Jha 2017, for a review) has been suggested as an outcome of the SD channel, indirectly through its young environment (Foley et al. 2009) and a good match of the SN properties to the prediction by a specific model of a failed/weak deflagration model (Kromer et al. 2013), and directly through the detection of a possible companion star for SN Iax 2012Z in a pre-SN image (McCully et al. 2014). The companion candidate of SN 2012Z is a blue object, and suggested to be a He star; it is thus not readily associated with the scenario proposed here. 

The fraction of SN1991T/1999aa-like bright SNe Ia in the volume limited sample account for $\sim 9$\% of the total SN Ia rate (Li et al. 2011). The over-luminous SNe Ia are rare, and its contribution to the total SN Ia rate is negligible as compared to the 1991T/1999aa-like SNe Ia. Therefore, at least $\sim 9$\% of the observed SNe Ia have a possible counterpart in the (H-accreting) SD evolution channel. This suggests that there could be a comparable number of `normal' SNe evolved through the `hidden' SD channel (i.e., Model high-B), which would however be identified preferentially as the DD channel in terms of natures of the CSM and companion star. Interestingly, our model predicts that for every potential SN Ia with distinct nature of the classical SD channel (Model non-B), there should be a similar number of normal-looking SNe Ia evolved through the SD channel under the magnetic field confinement (Model high-B). As a combination, the contribution of the SD channel could be non-negligible even in normal SNe Ia. Note however that addressing the properties of the highly-magnetized progenitor WD just before the explosions, and thus the nature of SNe Ia in Model high-B, requires further, detailed investigations.

\section{Summary}

The evolution of an accreting WD with strong magnetic field may differ from the classical SD channel (Ablimit \& Maeda 2019); the WD could steadily accrete the material from a donor to the Chandrasekhar mass even at a low mass transfer rate which would instead result in nova eruptions in the classical SD model. As a result, the parameter space for the SN Ia progenitors is different. Less massive donors spanning a larger range of the orbital period lead to the SN Ia explosion than in the classical SD channel. 

In this paper, we have focused on the WD/MS systems and performed the BPS simulations. The DTD from the accreting magnetic WD evolution toward the SN Ia progenitors spans a larger range of the delay times than the classical SD scenario; it ranges from $\sim 0.06$ Gyr to $\sim 11$ Gyr. Under the reasonable assumption that a fraction of the highly magnetic WD is $\sim 15$\%, the resulting DTD roughly follows the $t^{-1}$ power-law distribution. As such, the contribution of the SD channel toward SNe Ia can be as large as $\sim 50$\% among the whole SN Ia population without a major tension in the shape of the resulting DTD. 

Within the (WD/MS) SD channel, the contribution from the highly magnetized WD has been estimated to be comparable to that from the classical, non-magnetized WD channel. According to the BPS study, the sum of these contributions reach to $\sim 30$\% of the observed Galactic SN Ia rate. Adding the WD/RG channel (i.e., Yungelson et al. 1995; L$\ddot{\rm u}$ et al. 2009), the SD channel therefore would easily explain at least a half of the total SN Ia rate. 

The nature of the progenitor systems in the magnetic SD channel would probably resemble more the DD products rather than the classical SD products. They are expected to be associated with faint companion stars and rarefied CS environment. Further investigation is required to address the detailed properties of the immediate progenitor WD, and thus the nature of the resulting SNe, but we speculate that they show a rather uniform nature in the immediate progenitor WD, and thus in the SN ejecta. As such, it is likely that they would not be identified as the `SD' candidates through various observational constraints. Therefore, for every (potentially peculiar) SN observationally associated with the SD channel, we expect a comparable number of the `hidden' SD population in the normal class.

\begin{acknowledgements}

This work was funded by the JSPS International Postdoctoral fellowship (P17022). The work was supported by JSPS KAKENHI (grant no. 17F17022, 17H02864, 18H04585 and 18H05223).
\end{acknowledgements}


Ablimit, I., Xu, X.-J \& Li, X.-D., 2014, ApJ, 780, 80

Ablimit, I., Maeda, K. \& Li, X.-D., 2016, ApJ, 826, 53

Ablimit, I., \& Maeda, K., 2018, ApJ, 866, 151

Ablimit, I., \& Maeda, K. 2019, ApJ, 871, 31

Aubourg, E., Tojeiro, R., Jimenez, R., et al., 2008. A\&A 492, 631

Botticella, M.T., Riello, M., Cappellaro, E., et al., 2008. A\&A 479, 49

Cao, Y., Kulkarni, S. R., Howell, D. A., et al., 2015, Nature, 521, 328

Cappellaro, E. \& Turatto, M., 1997, in Proc. NATO Advanced Study Insti-
tute 486, Thermonuclear Supernovae, ed. P. Ruiz-Lapuente, R. Cannal, J.
Isern (Dordrecht: Kluwer), 77

Chomiuk, L., Soderberg, A. M., Moe, M., et al., 2012, ApJ, 750, 164

Cumming, A. 2002, MNRAS, 333, 589

Cropper M. 1990, Space Sci. Rew., 54, 195

Claeys, J. S. W., Pols, O. R., Izzard, R. G., Vink, J. \& Verbunt, F. W. M., 2014, A\&A, 563, 83

Das, U., Mukhopadhyay, B. \& Rao, A. R., 2013, ApJL, 767, L14

De, K., Kasliwal, M. M., Polin, A., et al., 2019, ApJ, 873, 18

Di Stefano, R., Voss, R. \& Claeys, J. S. W. 2011, ApJL, 738, L1

Dilday, B., Howell, D. A., Cenko, S. B., et al., 2012, Science, 337, 942

Farihi, J., Fossati, L., Wheatley, B. D. et al., 2017, arXiv:1709.08206v2

Ferrario, L., de Martino, D. \& Gnsicke, B. T., 2015, SSRv, 191, 111F

Foley, R. J.; Chornock, R., Filippenko, A. V., et al., 2009, AJ, 138, 376

Gonz$\acute{\rm a}$lez Hern$\acute{\rm a}$ndez, J. I., Ruiz-Lapuente, P., Tabernero, H. M., et al., 2012, Nature, 489, 533

Greggio, L. \& Renzini, A., 1983, A\&A, 118, 217

Hachisu, I., Kato, M., Saio, H. \& Nomoto, K., 2012, ApJ, 744, 69

Hachisu, I., Kato, M. \& Nomoto, K., 2012, ApJ, 756, 4

Hillebrandt, W. \& Niemeyer, J. C., 2000, ARA\&A, 38, 191

Holey, F. \& Folwer, W. A., 1960, ApJ, 132, 565

Hurley, Jarrod R., Tout, Christopher A. \& Pols, Onno R., 2002, MNRAS, 329, 897

Iben, I. \& Tutukov, A. V., 1984, ApJS, 54, 335

Ilkov, M., \& Soker, N. 2011, arXiv:1106.2027

Jha, Saurabh W., 2017, Handbook of Supernovae, Springer International Publishing AG, arxiv:1707.01110

Jiang, Ji-An, Doi, M., Maeda, K., et al., 2017, Nature, 550, 80

Jorgensen, H. E., Lipunov, V. M., Panchenko, I. E., et al., 1997, ApJ, 486, 110

Kahabka, P. 1995, ASP Conference Series, Vol. 85

 Kasen, D., 2010, ApJ, 708, 1025

Kelly, P. L., Fox, O. D., Filippenko, A. V., et al., 2014, ApJ, 790, 3

Kepler, S. O., Pelisoli, I., Jordan, S., Kleinman, S.J., et al., 2013, MNRAS, 429, 2934 

Kepler, S.O., Pelisoli, I., Koester, D., Ourique, G., et al., 2015, MNRAS, 446, 4078 

Kobayashi, C. \& Nomoto, K., 2009. ApJ 707, 1466

Kobayashi, C., Leung, S. C. \& Nomoto, K., 2019, arXiv:1906.09980

Kromer, M., Fink, M., Stanishev, V., et al., 2013, MNRAS, 429, 2287

Kroupa, P., Tout, Christopher A. \& Gilmore, G., 1993, MNRAS, 262, 545

Leloudas, G., Hsiao, E. Y., Johansson, J., et al., 2015, A\&A, 574, 61

Li, X.-D., \& van den Heuvel, E. P. J. 1997, A\&A, 322, L9

Li, W., Bloom, J.S., Podsiadlowski, P., et al., 2011, Nature, 480, 348

Li, W., Leaman, J., Chornock, R., et al., 2011, MNRAS, 412, 1441

Livio, M. 1983, A\&A, 121, L7

Livio, M. \& Mazzali, P., 2018, Phys. Rep., 736, 1

L$\ddot{\rm u}$, G., Zhu, C., Wang, Z., Wang, N., 2009, MNRAS, 396, 1086

Maeda, K., Kutsuna, M., Shigeyama, T., et al., 2014, ApJ, 794, 37

Maeda, K., Nozawa, T., Nagao, T. \& Motohara, K., 2015, MNRAS, 452, 3281

Maeda, K. \& Terada, Y. 2016, IJMPD, 253002

Maeda, K., Jiang, Ji-an, Shigeyama, T. \& Doi, M., 2018, ApJ, 861,78

Mannucci, F., Della Valle, M. \& Panagia, N., 2006. MNRAS, 370, 773

Mannucci, F., Maoz, D., Sharon, K., Botticella, M. T., Della Valle, M., Gal-Yam, A., \& Panagia, N. 2008, MNRAS, 383, 1121

Maoz, D. \& Badenes, C., 2010, MNRAS, 407, 1314

Maoz, D. \& Mannucci, F., 2012, PASA, 29, 447

Maoz, D., Mannucci, F. \& Nelemans, G., 2014, ARA\&A, 52, 107

Margutti, R., Soderberg, A. M., Chomiuk, L., et al., 2012, ApJ, 751, 134

Marietta, E., Burrows, A., Fryxell, B., 2000, ApJS, 128, 615

Mattila, S., Lundqvist, P., Sollerman, J., et al., 2005, A\&A, 443, 649

McCully, C., Jha, Saurabh W., Foley, Ryan J., et al., 2014, Nature, 512, 54

Nagao, T., Maeda, K. \& Yamanaka, M., 2017, ApJ, 835, 143

Noebauer, U. M., Kromer, M., Taubenberger, S., et al., 2017, MNRAS, 472, 2787

Nomoto, K., 1982, ApJ, 257, 780

Osborne, et al. 2001, A\&A, 378, 800

Pala, A. F., G$\ddot{a}$nsicke, B. T., Breedt, E., et al., 2019, arXiv:1907.13152

Riess, A. et al., 1998, AJ, 116, 1009

Ruiz-Lapuente, P., 2014, New Astron. Rev., 62, 15

Schaefer, B. E. \& Pagnotta, A., Nature, 2012, 481, 164

Schawinski, K., 2009, MNRAS, 397, 717

Schmidt, G. D., Hoard, D. W., Szkody, P., Melia, F., Honeycutt, R. K. \& Wagner, R. M., 1999, ApJ, 525, 407

Shapiro, S. L. \& Teukolsky, S. N., Black holes, white dwarfs, and neutron stars: The physics of compact objects., Research supported by the National Science Foundation. New York, Wiley-Interscience, 1983, 663 p.

Stritzinger, M. D., Shappee, B. J., Piro, A. L., et al., 2018, ApJ 864, 35

Soker, N. 2019, arXiv:1905.06025

Sokoloski, J. L. \& Beldstin, L., 1999, ApJ, 517, 919

Tucker, M. A., Shappee, B. J. \& Wisniewski, J. P., 2019, ApJ, 872, 22

Totani, T., Morokuma, T., Oda, T., Doi, M. \& Yasuda, N., 2008, PASJ 60,
1327

Tutukov, A.V. \& Yungelson, L.R., 1979, Acta Astron., 29, 665

Uenishi, T., Nomoto, K. \& Hachisu, I., 2003, ApJ, 595, 109

Wang, B. \& Han, Z., 2009. A\&A 508, L27

Wang, B. \& Han, Z. 2012, New Astron. Rev., 56, 122

Wang, B., 2018, RAA, 18, 49

Wang, C., Jia, K. \& Li, X.-D., 2016, MNRAS, 457, 1015

Webbink, R. F., 1984, ApJ, 277, 355

Wheeler, J. C. 2012, ApJ, 758, 123

Whelan, J. \& Iben, I., 1973. ApJ 186, 1007

Willems, B. \& Kolb, U., 2004, A\&A, 419, 1057

Yamanaka, M., Maeda, K., Tanaka, M., et al., 2016, PASJ, 68, 68

Yungelson, L. R., Livio, M., Tutukov, A. V., \& Kenyon, S. J. 1995, ApJ, 477, 656

Yoon, S. -C. \& Langer, N., 2004, A\&A, 419, 645

Yungelson, L.R. \& Livio, M., 2000. ApJ 528, 108


\begin{figure}
\centering
\includegraphics[totalheight=4.0in,width=5.0in]{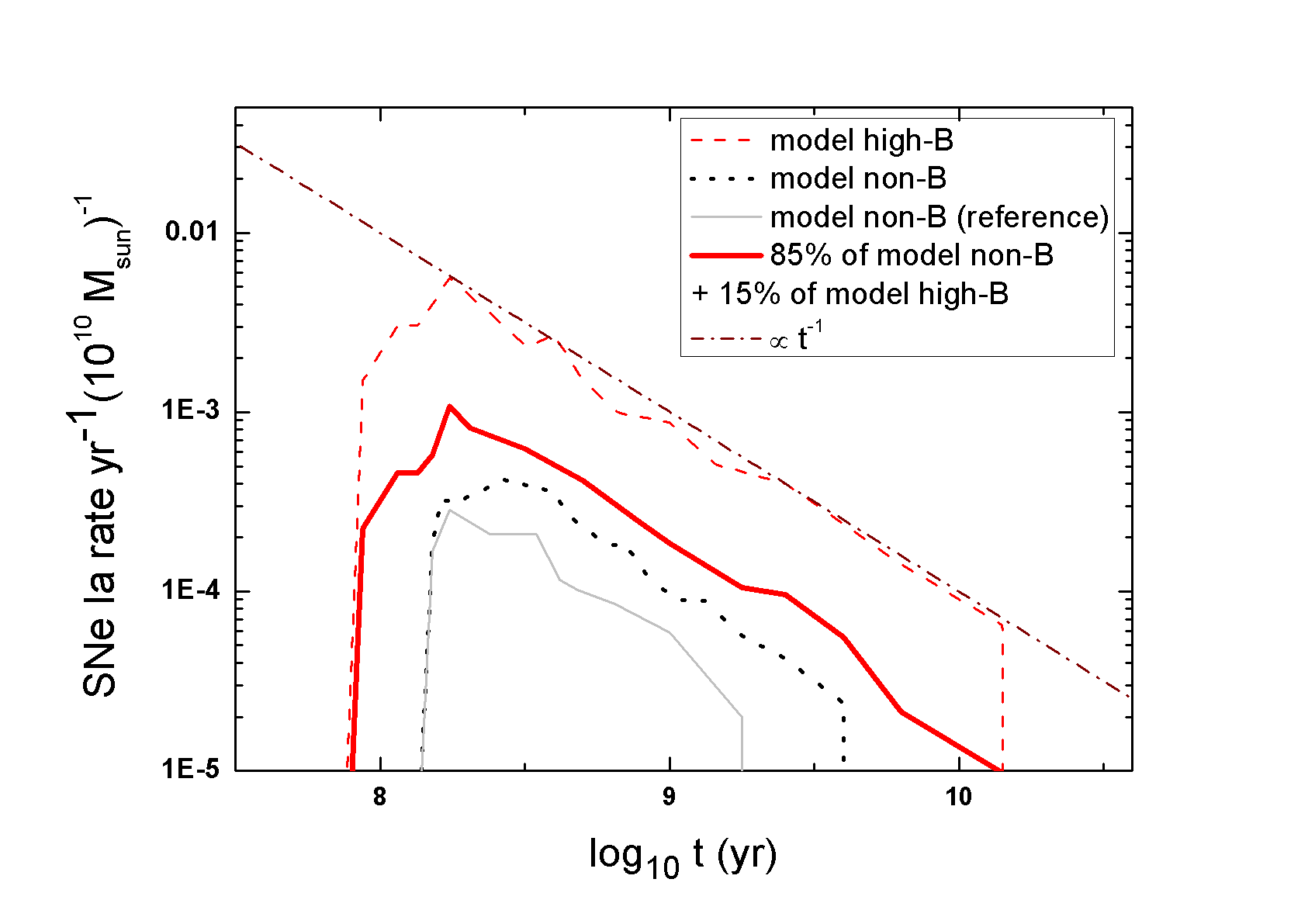}
\caption{Delay time distributions (DTDs) from the BPS simulations. The DTDs from Models non-B and high-B are shown by the black dotted curve and the thin dashed red curve, respectively. The combination of these two models, with relative fractions of 85\% and 15\%, is shown by the thick red curve. As a reference, Model non-B with the binding energy parameter set to be $\lambda_{\rm g}$ is shown by the gray curve. The wine dash--dotted line shows the $\sim t^{-1}$ power-law distribution as inferred from observations.}
\label{fig:1}
\end{figure}

\clearpage

\begin{figure}
\centering
\includegraphics[totalheight=4.0in,width=5.0in]{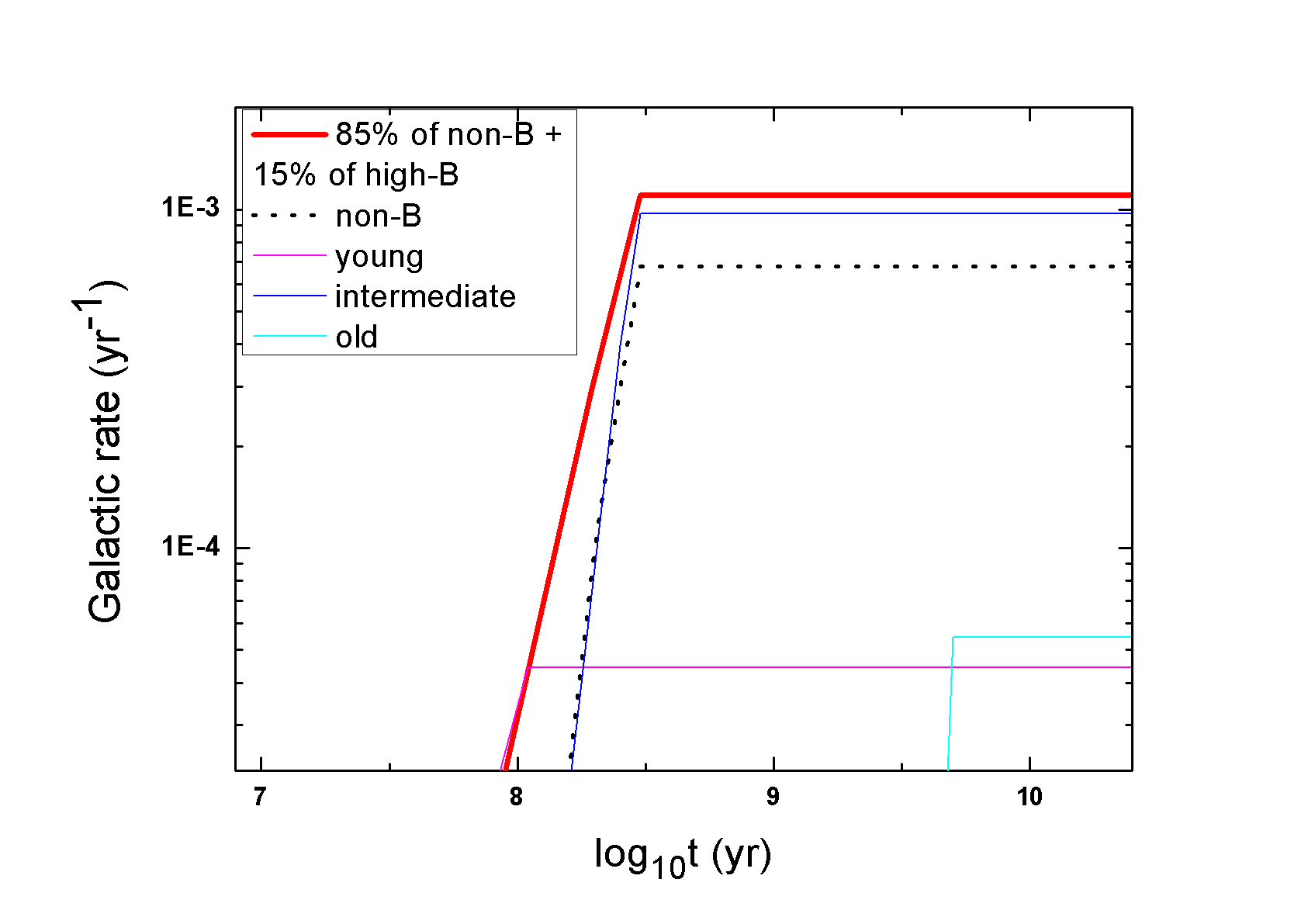}
\caption{Evolution of the Galactic birthrates of the Chandrasekhar C+O WD (i.e., the SN Ia progenitor) within the WD/MS binaries. A constant star formation rate is assumed. The black dotted line shows the classical WD/MS channel (i.e., Model non-B). The rate calculated by a combination of Models high-B (15\%) and non-B (85\%) is shown by the thick red line; this model is further divided into young ($\rm{log_{10}}\,t<8.0$; magenta line), intermediate ($8.0\leq \rm{log_{10}}\,t<9.6$; blue), and old ($\rm{log_{10}}\,t \geq 9.6$; cyan) populations. }
\label{fig:2}
\end{figure}

\clearpage

\begin{figure}
\centering
\includegraphics[totalheight=3.6in,width=4.2in]{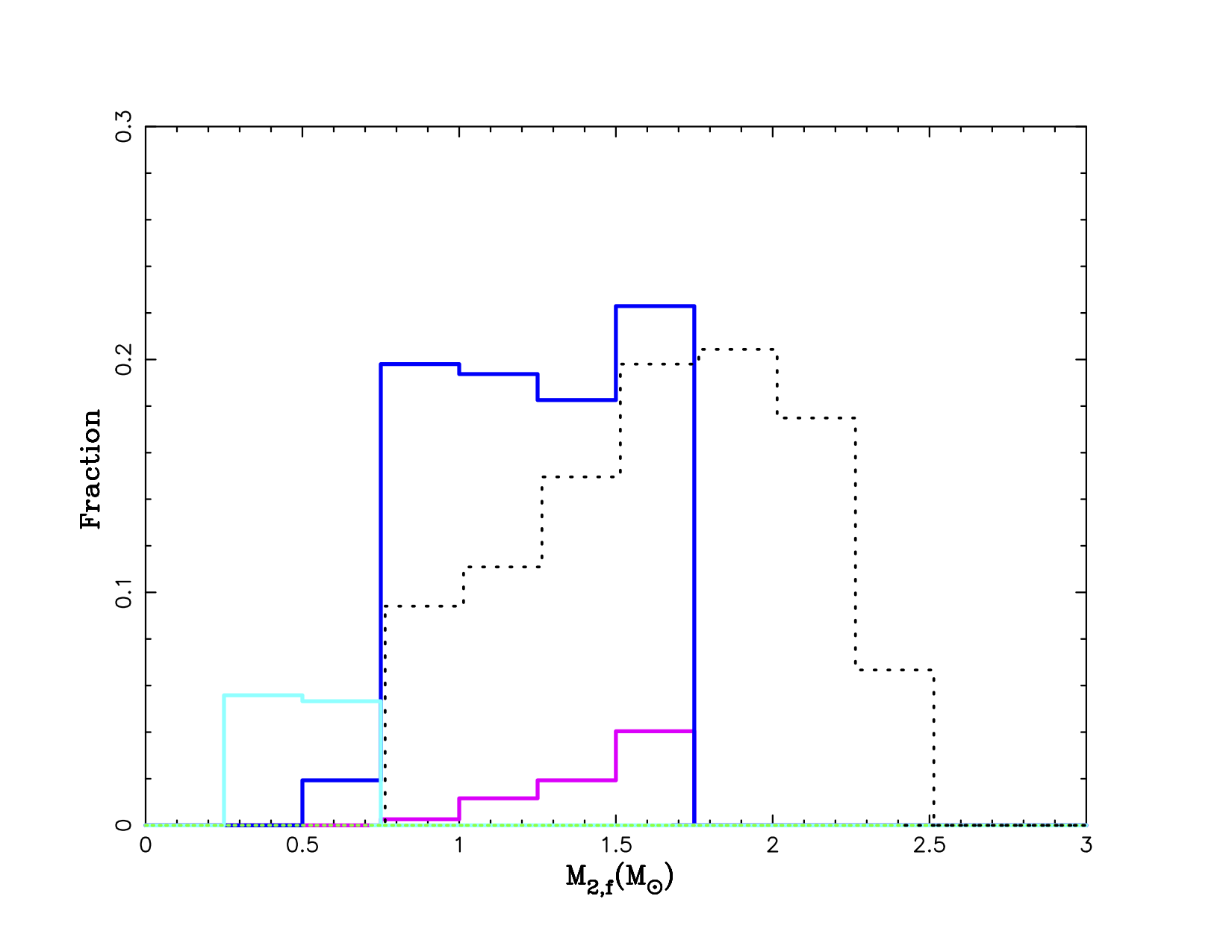}
\includegraphics[totalheight=3.6in,width=4.2in]{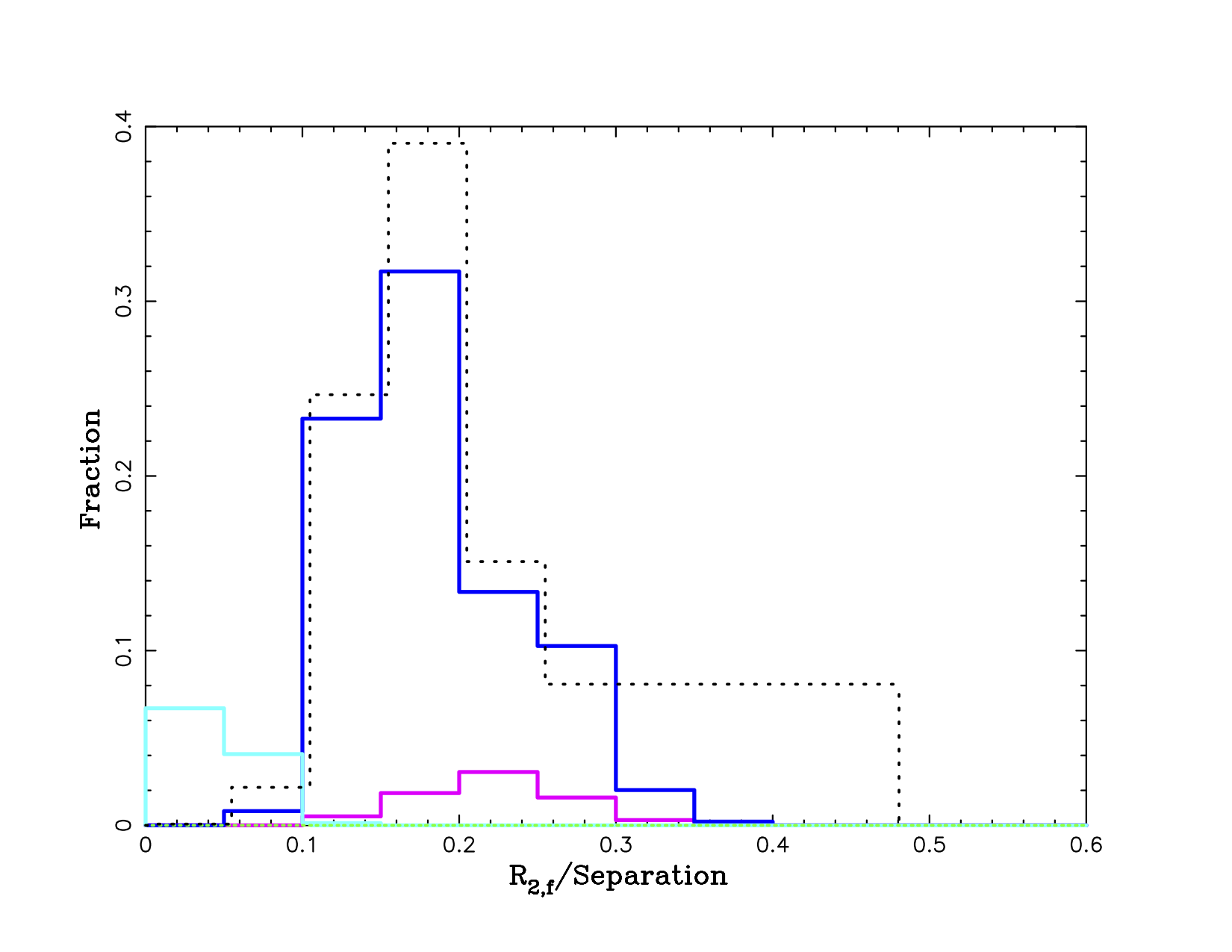}
\caption{Final distributions of the companion masses, and the companion radii relative of the binary separation. The distribution is separately normalized for Model non-B and Model high-B. The black dotted line shows Model non-B. The solid lines are for Model high-B, divided into the young, intermediate and old populations (same as Figure 2). Note that all the SNe Ia progenitors from Model non-B belong to the intermediate population. 
}
\label{fig:3}
\end{figure}

\clearpage

\end{document}